\documentclass[aps,reprint,graphicx]{revtex4-1}

\usepackage{graphicx}
\usepackage{epstopdf}
\usepackage{dcolumn}
\usepackage{bm}
\usepackage{amsmath}
\usepackage{lipsum}

\newcommand{\fref}[1]{Figure~\ref{#1}}
\newcommand{\sref}[1]{Section~\ref{#1}}

\draft 

\begin{document}

\title{Modeling hard-soft block copolymers as a liquid crystalline polymer}

\author{M.~Manav}
\email{mmanav@ethz.ch}
\affiliation{Mechanical and Process Engineering, ETH Zurich, Switzerland}

\author{M.~Ponga}
\email{mponga@mech.ubc.ca}
\affiliation{Mechanical Engineering, University of British Columbia, Vancouver, Canada}

\author{M.~Ortiz}
\email{ortiz@caltech.edu}
\affiliation{California Institute of Technology, Pasadena, USA}

\date{\today}

\begin{abstract}
We report a new computational approach to model hard-soft block copolymers like polyurea as a liquid crystalline polymer to understand their microstructural evolution due to mechanical loading. The resulting microstructure closely resembles the microstructure observed in polyurea. The stress-strain relations in uniaxial compression and tension loading obtained from the model are also in close quantitative agreement with the experimental data for polyurea. We use the model to elucidate the evolution of the hard and the soft domains during loading, which is consistent with the experimental measurements characterizing microstructural evolution in polyurea.

\end{abstract}

\maketitle 

%

\section{Introduction}
\label{intro}
Thermoplastic elastomers (TPEs) such as polyurea~\cite{primeaux2004polyurea} exhibit excellent mechanical strength and toughness as well as resistance to shock and impact loading~\cite{roland2007high,bogoslovov2007impact,jiao2014measurement,mott2016deformation}. These properties have enabled the application of polyurea in bulletproof vests and helmets~\cite{walsh2005development,grujicic2010blast}, blast mitigation~\cite{porter2002polymer,bahei2006blast,tekalur2008blast}, among others, and have led to a surge in scientific interest in modeling it and understanding the microstructural origins of its mechanical response~\cite{mott2016deformation,iqbal2016polyurea,zheng2022molecular}. These remarkable mechanical properties of polyurea are often attributed to its hierarchical microstructure. A polyurea molecule consists of alternate blocks of flexible aliphatic segments and stiff segments made of aromatic moieties. In polyureas with long flexible segments, the two blocks phase-segregate forming ribbon-shaped hard domains ($\approx 5-10$ nm wide with  $\approx 7$ nm separation in polyurea P1000) made of stiff segments dispersed in a soft matrix of flexible blocks~\cite{pathak2008structure, rinaldi2011microstructure, grujicic2012experimental, castagna2012role, choi2012microstructure, castagna2013effect}. The hard domains are glassy due to the hydrogen bond network and $\pi-\pi$ stacking of the neighboring aromatic moieties. They act as physical crosslinks as well as stiffeners. However, the hierarchical arrangement also makes it challenging to understand the microstructural origin of its properties.

A number of experimental studies have been conducted to characterize the evolution of polyurea microstructure due to mechanical loading. Wide angle X-ray diffraction (WAXD) studies have shown that the mean spacing between hard segments does not change during quasi-static uniaxial tensile loading, though it leads to their alignment along tensile direction~\cite{rinaldi2011microstructure,choi2012microstructure}. The alignment is strain-rate dependent and observed to be absent at high strain-rates~\cite{pathak2008structure}. Small angle X-ray scattering (SAXS) measurements reveal an axial increase and a transversal decrease in the interdomain spacing of hard domains with increasing strain~\cite{rinaldi2011microstructure}. They also indicate a breakdown in the hard domain network structure at true strains above $\sim 0.4$. In situ AFM tapping-phase study of polyurea also has revealed fragmentation of hard domains upon rapid loading and various nano and mesoscale phase transitions upon relaxation under fixed tensile strains~\cite{jin2022understanding}. A study of the microstructure of polyurea under hydrostatic compression in a diamond anvil cell using WAXD and SAXS revealed the presence of interchain hydrogen-bonded network even at pressures well above 1 GPa though crystallinity is not observed~\cite{rosenbloom2021microstructural}. The study also reported phase-mixing with increasing pressure and pointed to the contrasting behavior of polyurea under hydrostatic compression and in tension where hard domains become oriented. Investigation of the atomic-scale morphology of the domains using multi-angle energy-dispersive X-ray diffraction at pressures up to 6 GPa shows reduced mean spacing between nearby hard segments in contrast to minimal deformation within a hard segment and a decrease in  interdomain order with increasing pressure~\cite{eastmond2021probing}.

Molecular dynamics (MD) simulation has also been employed to investigate the mechanical behavior of polyurea. An all-atom (AA) model of polyurea leveraging its phase-segregated microstructure and performing molecular dynamics simulation of each domain independently before combining their properties like in a composite yielded close agreement with their experimentally measured quasistatic response \cite{heyden2016all} and shock response~\cite{manav2021}. Molecular scale deformation mechanisms in polyurea during mechanical loading are investigated using AA MD simulation in~\cite{roy2022investigating}. Shock response of polyurea has also been investigated using AA MD simulation~\cite{grujicic2012molecular,grujicic2015experimental,dewapriya2022quantum} and inelastic deformation, hydrogen bond breaking and reforming, and permanent densification have been identified as contributors to shock energy dissipation. These models can typically access length and time scales only up to the order of $10~{\rm nm}$ and  $10~{\rm ns}$, respectively, which is insufficient to capture the evolution of polyurea microstructure. To elucidate the effect of the phase-segregated microstructure, coarse-grained models~\cite{agrawal2016prediction,yao2020mesoscale} and generic bead-spring polymer models with control over the strength of pair-interaction potential to simulate hard and soft segments~\cite{arman2012viscoelastic,cui2013thermomechanical,zhang2018mechanical} have also been reported. They suggest that the multiblock architecture of polyurea leading to the formation of interconnected hard domains results in improved mechanical properties. A hybrid all-atom coarse-grained model was employed to study the toughness of polyurea, and self-reinforcement of oriented, hard segments and stress-dependent release of soft segments from hard domains are identified as two key mechanisms~\cite{zheng2022molecular}.

\begin{figure*}
	\centering
	\includegraphics[width=0.75\linewidth]{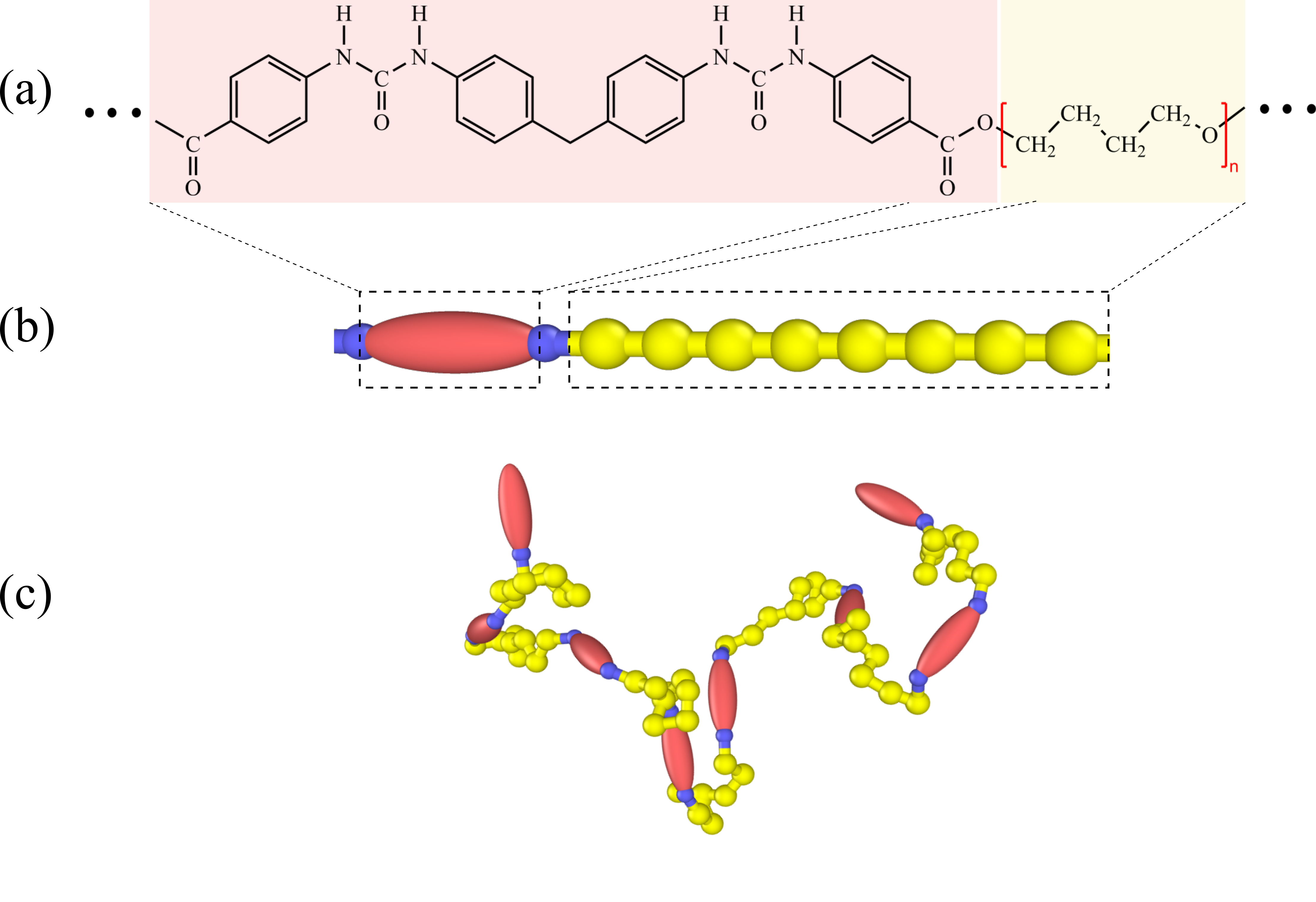}
	\caption{(a) A repeating unit of a polyurea molecule. $n=14$ for polyurea P1000 used for comparison in this work. (b) A repeating unit of a coarse-grained molecule, and (c) a coarse-grained molecule used in the simulation. Stiff blocks with aromatic moieties in polyurea have been replaced with ellipsoidal particles in the coarse-grained molecule and flexible blocks with point particles. To connect the point particles to the ends of the ellipsoidal particles, a dummy point particle with negligible mass is first rigidly attached to the ends of ellipsoids. Adjacent point particles, point particles, and dummy particles are connected by bonds.}
	\label{fig1}
\end{figure*}

We report here a generic coarse-grained model of polyurea with the aim of elucidating the evolution of their microstructure in response to mechanical loading. In a departure from the earlier models, we leverage the rigidity and the shape anisotropy of the hard segments in a polyurea molecule to model them as ellipsoidal particles like in a liquid crystalline polymer~(\fref{fig1}). Since liquid crystals exhibit a complex phase diagram in temperature-pressure-density space~\cite{de2002global}, this modeling approach also facilitates an investigation of the influence of their phase transitions on the mechanical response of the modeled polymer. We show that the model produces ribbon-like hard domains resembling that in polyurea and the stress-strain relation obtained from the model is in close agreement with the experimental data for polyurea P1000 (henceforth referred to as polyurea) for which the number of repeating units in soft segments is $n=14$ (see \fref{fig1}(a)). We then study its microstructural evolution during uniaxial loading.

The paper is organized as follows. \sref{method} details the model and methodology. \sref{result} presents and discusses results from the model, including stress-strain relation and microstructural evolution with mechanical loading. The paper ends with conclusions and future directions in \sref{conclusions}.

\section{Molecular dynamics simulation and analysis}
\label{method}
We have developed a generic coarse-grained model of a molecule as shown in \fref{fig1} and performed molecular dynamics simulation using LAMMPS~\cite{LAMMPS}. Lennard-Jones (LJ) units have been used throughout this paper, where $m,~\sigma$ and $\epsilon$ are mass, length, and energy scales, respectively. The time scale is denoted by the characteristic time $\tau$. We define a coarse-grained molecule by connecting ellipsoidal particles and points particles (\fref{fig1}). To create a bonded connection between a point particle and \emph{ends} of an ellipsoidal particle, we rigidly attached \emph{dummy} point particles of very small mass to the ends of an ellipsoidal particle. Then, point particles were connected to these dummy particles using bonded interaction. Each molecule consists of eight ellipsoids connected with a chain of eight point particles in between two consecutive ellipsoids. The number of point particles is chosen to ensure that a phase-segregated microstructure forms when a melt of the polymer is cooled~\cite{zhang2018mechanical}. The diameter of an ellipsoidal particle is $3\sigma$, $\sigma$, and $\sigma$ along its three principal axes. The masses of an ellipsoidal particle, a point particle, and a dummy particle are $1.5708m$, $0.5236m$, and $0.0005m$, respectively. 

The total potential energy consists of pair energy ($\mathcal{E}_{pair}$) and bond energy ($\mathcal{E}_{bond}$).
\begin{align}
	&\mathcal{E}=\mathcal{E}_{pair}+\mathcal{E}_{bond}, \\
	&\mathcal{E}_{pair} = \sum_{j>i}^{i, j \in {\rm point~particles}}U_{ij}^{LJ}+ \sum_{l>k}^{k\lor l \in {\rm ellipsoids}}U_{kl}^{GB}.
	\label{eq1}
\end{align}
$\mathcal{E}_{pair}$ accounts for the interactions between a pair of point particles, a pair of ellipsoidal particles, and an ellipsoidal and a point particle. $\mathcal{E}_{bond}$ is the energy associated with bonded interaction.

In our model, nonbonded point particles interact with a Lennard-Jones (LJ) potential, given as follows:
\begin{equation}
	U_{ij}^{LJ} = 4\epsilon\left(\left(\frac{\sigma}{r_{ij}}\right)^{12}-\left(\frac{\sigma}{r_{ij}}\right)^{6}\right) - U_{shift},
\end{equation}
where $\epsilon$, and $\sigma$ are characteristic energy and length. The potential is cut-off at a distance $r_c=2.5\sigma$ and shifted by $U_{shift}$ to ensure its continuity. In the simulations, we set $\epsilon=1$ and $\sigma=1$.

Interaction of an ellipsoidal particle with other particles is governed by Gay-Berne (GB) potential $U_{kl}^{GB}$, which is a product of three terms, following~\cite{berardi1998gay,everaers2003interaction,Brown09}:
\begin{equation}
	U_{kl}^{GB} = U_r\cdot \eta \cdot \chi,
	\label{eq3}
\end{equation}
where $U_r$ is a shifted LJ interaction, and distance-independent interaction anisotropy is characterized by $\eta$ and $\chi$. $\eta$ and $\chi$ control the interaction strength based on particle shapes and relative potential well depths, respectively. Shape for a particle $i$ is specified by the shape matrix ${\mathbf S_i} = {\rm diag}(a_i,~b_i,~c_i)$, where $a_i,~b_i$ and $c_i$ are ellipsoidal semiaxes. Orientation of particle $i$ is given by ${\mathbf A_i}$ which represents the rotational transformation from the simulation cell frame to the body frame. Relative energy matrix is given by ${\mathbf E_i} = {\rm diag}(\epsilon_{ai},~\epsilon_{bi},~\epsilon_{ci})$, where $\epsilon_{ai},~\epsilon_{bi}$ and $\epsilon_{ci}$ are relative well depths. Then, $U_r$ is given by:
\begin{equation}
	U_r = 4k_{\epsilon}\epsilon(\rho^{12} - \rho^6), \\
	\rho = \frac{\sigma}{h + \gamma\sigma},
	\label{eq4}
\end{equation}
where $\epsilon$ and $\sigma$ are the same as in LJ potential, $k_{\epsilon}$ is used to tune the depth of the potential well, $\gamma$ is the shift parameter and $h$ is the distance of closest approach of interacting particles and is approximated as:
\begin{equation}
	h = r - \left(\frac{1}{2}\hat{{\mathbf r}}^T{\rm\bf G^{-1}}\hat{{\mathbf r}}\right)^{-1/2},
	\label{eq5}
\end{equation}
where ${\rm\bf r}={\rm\bf r}_l-{\rm\bf r}_k$ is the vector between centers of ellipsoids $k$ and $l$ with magnitude $r = \sqrt{\rm\bf r\cdot \rm\bf r}$ and unit vector $\hat{{\rm\bf r}} = {\rm\bf r}/r$. Also, ${\rm\bf G} = {\rm\bf A_k^T}{\rm\bf S_k^2}{\rm\bf A_k} + {\rm\bf A_l^T}{\rm\bf S_l^2}{\rm\bf A_l}$.

$\eta$ is given as follows:
\begin{equation}
	\eta = \left( \frac{2s_k s_l}{\rm det(\bf G)} \right)^{\upsilon/2},
	\label{eq6}
\end{equation}
where $s_i = \beta_i(a_i^2 +b_i c_i)(b_i c_i)^{1/2}$ for particle $i$, and $\chi$ is given by the following expression:
\begin{equation}
	\chi = (2\hat{{\mathbf r}}^T{\rm\bf B^{-1}}\hat{{\mathbf r}})^{\mu},
	\label{eq8}
\end{equation}
where ${\rm\bf B} = {\rm\bf A_k^T}{\rm\bf E_k^{-1/\mu}}{\rm\bf A_k} + {\rm\bf A_l^T}{\rm\bf E_l^{-1/\mu}}{\rm\bf A_l}$. $\upsilon$ and $\mu$ are empirical exponents used to tune the potential. The formulation in~\cite{berardi1998gay}, which is implemented in LAMMPS, assumes that the longest ellipsoidal semiaxis is the third semiaxis $c_i$ as also pointed out in~\cite{everaers2003interaction}. $\beta_i$ is introduced here as a correction factor if that is not the case.

In our simulation, parameter values for ellipsoids are chosen to be the classical values~\cite{gay1981modification} for which phase-diagram has been studied extensively~\cite{de2002global}. We take $a_i=1.5\sigma$ and $b_i=c_i=0.5\sigma$. Also, $\epsilon_{ai}=0.2$ and $\epsilon_{bi}=\epsilon_{ci}=1$, where the values are for particles interacting along end-to-end, side-to-side, and face-to-face semiaxes~\cite{berardi1995generalized}, respectively. We also take $\gamma=1$, $\upsilon=1$, and $\mu=2$. The selected cut-off for ellipsoidal particles is $4\sigma$. Furthermore, since $a_i$ is taken as the longest semiaxis for ellipsoid $i$, the correction factor is $\beta_i=4\sqrt{3}/10$ for the prescribed values of semiaxes.

Bonded interaction is governed by the Finitely extensible nonlinear elastic (FENE) potential as given below.
\begin{align}
	U_b = &-\frac{1}{2}KR_0^2\log\left(1-\left(\frac{r}{R_0}\right)^2 \right) + \nonumber \\ 
	&4\epsilon\left(\left(\frac{\sigma}{r} \right)^{12}-\left(\frac{\sigma}{r} \right)^6 \right)+\epsilon,
	\label{eq9}
\end{align}
In the FENE potential $r$ represents the  bond length, and $R_0=1.5\sigma$ is the maximum bond length. At initial elevated temperatures during simulation to prepare a well mixed melt, $K=20T^*\epsilon/\sigma^2$ was employed to control maximum stretch in bonds, where $T^*$ is reduced temperature and is given by $T^*=T/\left(\frac{\epsilon}{k_B}\right)$. Once the melt was cooled down to $T^*=1.5$, we set $K=30\epsilon/\sigma^2$.
We employed time-reversible integrators presented in~\cite{kamberaj2005time} to evolve the translational and rotational motion of the rigid bodies consisting of an ellipsoid and dummy particles, and velocity-Verlet scheme for the translational motion of point particles. Ellipsoidal particles have only five degrees of freedom (DOF) since they are not free to spin around their longest principal axis. So, we block this DOF in our simulation and remove it from the calculation of temperature. We used the Nos\'e-Hoover thermostat for point particles. Chain thermostat and barostat based on generalized Nos\'e-Hoover approach~\cite{martyna1992nose} were used to control the temperature of the rigid bodies and the pressure in the simulation cell. Note that while barostating is based on the rigid bodies, the positions of all the particles in the simulation cell are updated. A timescale of $0.5\tau$ was used for the thermostat and $5\tau$ was used for the barostat. Throughout the simulation, a time step of $0.005\tau$ was employed.

\subsection{Representative volume element (RVE)}
To obtain glassy hard domains, we developed RVEs for $k_{\epsilon}=2,~3$ and $5$, ensuring stronger interaction between ellipsoids compared to the interaction between point particles and following~\cite{cui2013thermomechanical,zhang2018mechanical}. In our simulation, we placed 1000 molecules in a simulation box in a random configuration at a high temperature to obtain a well mixed melt. The initial melt was equilibrated at a high temperature and a small pressure, before being cooled down to $T^* = 1.5$ and releasing the pressure. Simulation cells become unstable at high temperatures in an NPT simulation due to the limited cut-off range of the pair potentials. The pressure was applied to keep the simulation cells stable. The raised temperature and pressure for $k_{\epsilon}=2$ were $T^* = 2$ and $p^*=0.1$ ($p^*$ is reduced pressure, $p^*=p/\left(\frac{\epsilon}{\sigma^3}\right)$). They were set to be $T^* = 2.5$ and $p^*=0.3$ for $k_{\epsilon}=3$, and $T^* = 3$ and $p^*=0.6$ for $k_{\epsilon}=5$. Then the cells were equilibrated at $T^* = 1.5$ for $10^6$ time steps. All the equilibrated melt at $T^* = 1.5$ were cooled to a reduced temperature of $T^* = 0.55$. For $k_{\epsilon}=5$, ellipsoidal particles phase separate even at $T^* = 1.5$. But for the lower values of $k_{\epsilon}$, we observed phase-segregated microstructure with ribbon-shaped hard domains at $T^* = 0.55$. The simulation cell at $T^* = 0.55$ was further equilibrated for at least $5\times10^6$ time steps before using it as an RVE for the production run. Since the glass transition temperature of a polymer melt of LJ particles with $\epsilon = 1$ and $\sigma=1$ is $\approx 0.45$~\cite{buchholz2002cooling,cui2013thermomechanical}, it is in a melt state at  $T^* = 0.55$. However, the ellipsoidal particles are in solid state, as indicated by their translational diffusion and rotational autocorrelation.

\subsection{Mean squared displacement and autocorrelation}
We calculate mean squared displacement ($MSD$) and rotational autocorrelation ($C_R$) for the ellipsoidal particles, which are given as follows:
\begin{align}
	&MSD(\Delta t) = \langle |{\mathbf r_i(t_0+\Delta t) - \mathbf r_i(t_0)}|^2\rangle,\\
	&C_R(\Delta t) = \langle {\mathbf u_i(t_0+\Delta t) \mathbf u_i(t_0)}\rangle,
	\label{eq11}
\end{align}
where $\Delta t$ is the time interval, $\mathbf r_i(t)$ and $\mathbf u_i(t)$ are the position and orientation vectors of the $i_{th}$ ellipsoid at time $t$, and $\langle \cdot\rangle$ is phase average.

\subsection{Microstructure characterization}
To characterize the hard domain, its radial distribution function (RDF) and orientational order parameter are calculated. The nematic tensor is defined as:
\begin{equation}
	{\mathbf Q} = \frac{1}{2}\langle 3{\mathbf u_i \mathbf u_i^T} - {\mathbf I}\rangle,
	\label{eq10}
\end{equation}
where ${\mathbf I}$ is identity matrix. The largest eigenvalue of ${\mathbf Q}$ is the order parameter $S$ and the corresponding eigenvector is the orientation of the director alignment. In uniaxial compression, directors align in the plane perpendicular to the direction of compression. Then the eigenvalue with a negative value has the largest magnitude and we take that as the order parameter.

To characterize the deformation of the soft domain, we investigate the mean and distribution of the end-to-end distance $R_{ee}$ of the soft segments connecting two ellipsoidal particles in a molecule.

\subsection{Uniaxial mechanical loading}
In uniaxial tension and compression tests, the sample is stretched (compressed) at a fixed engineering strain rate along $x$-direction. In $y$ and $z$-directions, pressure is maintained at zero. We have performed tests at three strain rates: $5\times 10^{-3}~\tau^{-1}$, $5\times 10^{-4}~\tau^{-1}$ and $2.5\times 10^{-5}~\tau^{-1}$.

\section{Result and discussion}
\label{result}
In this section, we present and discuss the results from the simulation, including initial configuration, mechanical response, and microstructure evolution due to the applied compressive and tensile uniaxial loading.

\subsection{Initial configuration}
\begin{figure*}
	\centering
	\includegraphics[width=0.85\linewidth]{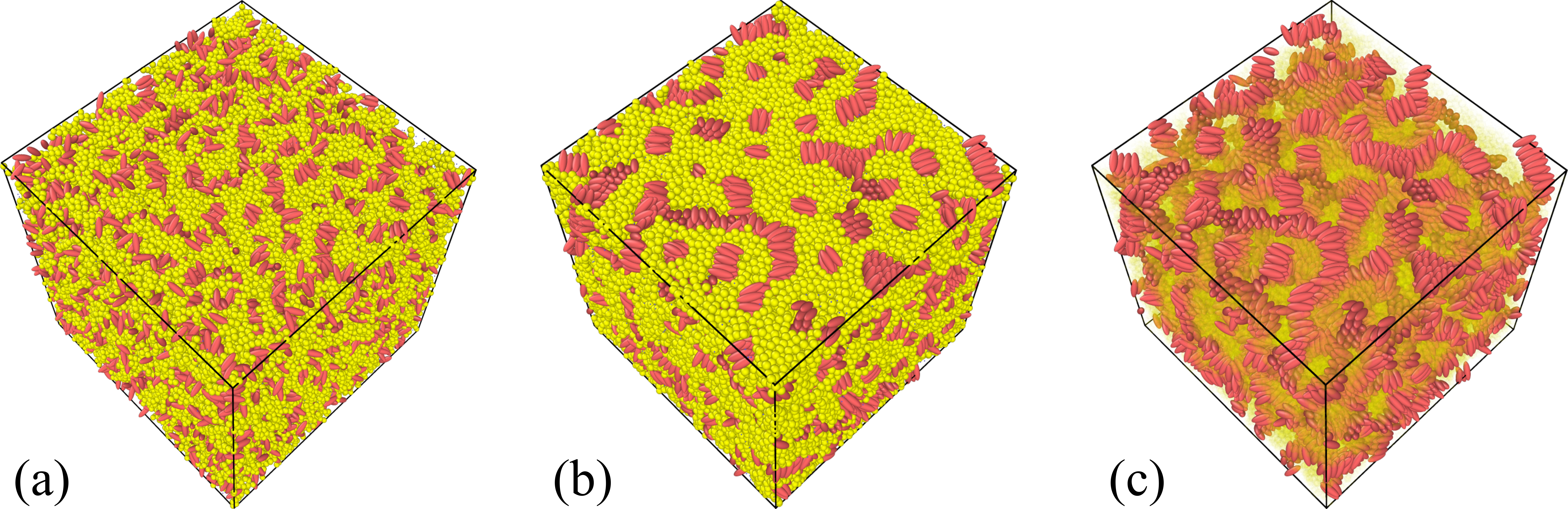}
	\caption{(a) Homogeneous polymer melt at $T^*=1.5$, (b) phase-separated RVE at $T^*=0.55$, and (c) same as (b) but with translucent soft domain exhibiting ribbon-shaped hard domain.}
	\label{fig2}
\end{figure*}
\fref{fig2}(a) shows a well-mixed and homogeneous melt at $T^*=1.5$ for $k_{\epsilon}=3$. It forms phase-segregated domains at $T^*=0.55$ as shown in \fref{fig2}(b). Also, the phase-segregated hard domains are ribbon-shaped as exhibited by \fref{fig2}(c) in close agreement with the phase-segregated domains in polyurea. Furthermore, even though the ribbon-shaped domains appear to be disconnected in \fref{fig2}(b), cluster analysis confirms that they are connected and form a scaffold as seen in \fref{fig2}(c).

\begin{figure}[!ht]
	\centering
	\includegraphics[width=0.9\linewidth]{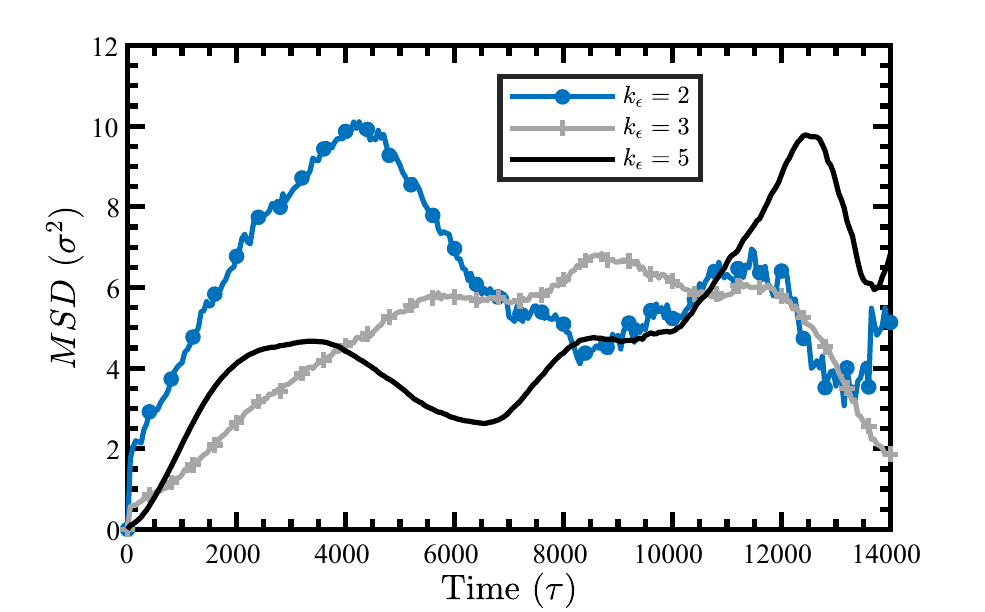}
	\includegraphics[width=0.9\linewidth]{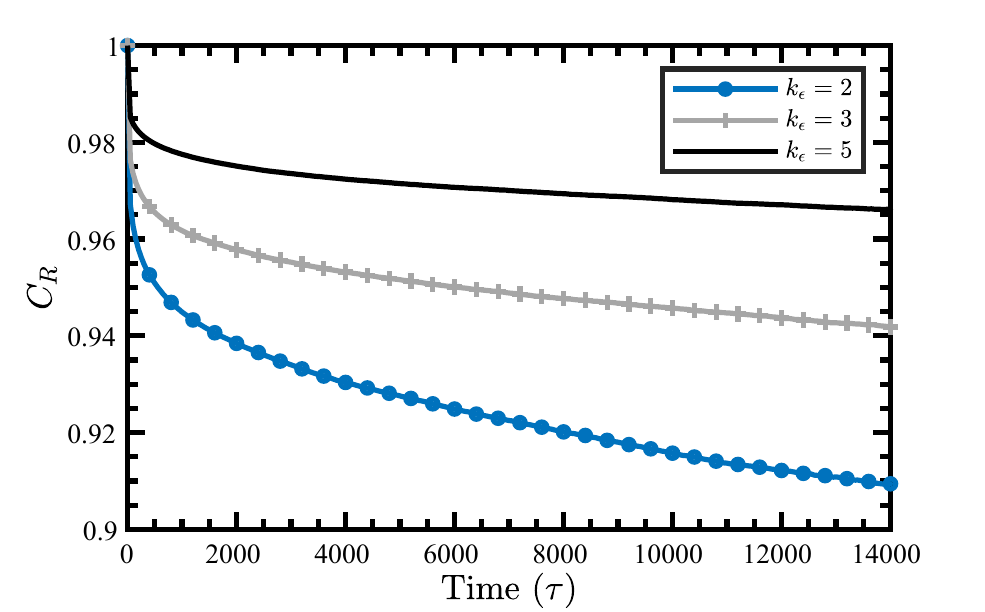}
	\caption{Evolution of the mean squared displacement (top), and the rotational autocorrelation (bottom) of the ellipsoidal particles with time in equilibrium condition at $T^*=0.55$.}
	\label{fig3}
\end{figure}
To elucidate the glassy nature of the hard domain, mean squared displacement and rotational autocorrelation of the ellipsoidal particles are shown in \fref{fig3}. The small magnitude of $MSD$ and very slow decay of $C_R$ over the simulation time, which is of the order of the inverse of the slowest applied strain rate in uniaxial tests, allude to their being in a glassy state. Also, a stronger interaction between the ellipsoidal particles (higher $k_{\epsilon}$) leads to a slower decay of $C_R$.

\subsection{Mechanical response}
\begin{figure}[!ht]
	\centering
	\includegraphics[width=0.9\linewidth]{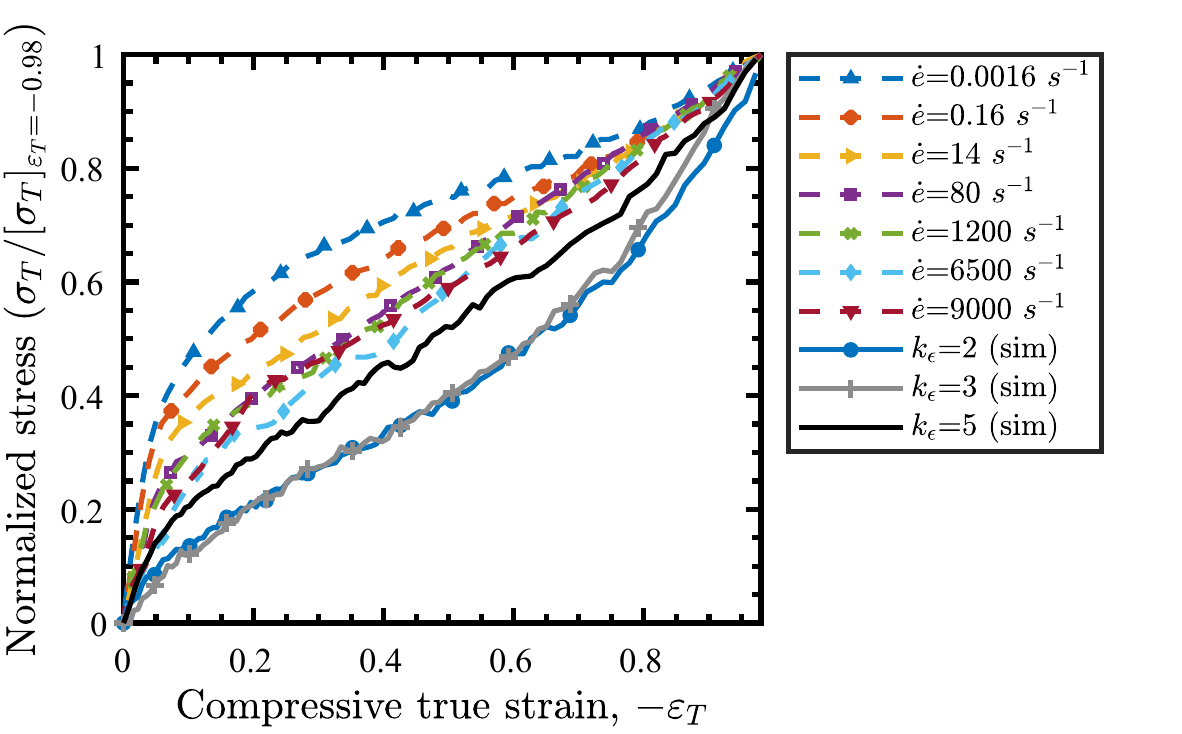}
	\includegraphics[width=0.9\linewidth]{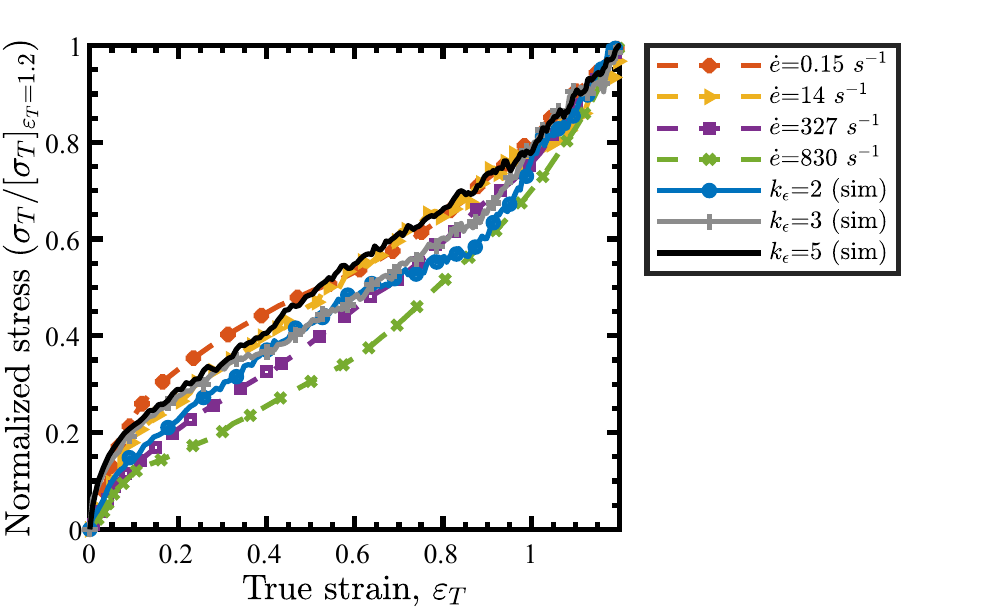}
	\caption{Uniaxial stress-strain curves in compression (top) and tension (bottom) at engineering strain rate $\dot{e}=5\times 10^{-5}~\tau^{-1}$ in simulation is compared with the experimental data for polyurea. Experimental data in compression and tension are taken from~\cite{sarva2007stress} and~\cite{roland2007high,pathak2008structure}, respectively.}
	\label{fig4}
\end{figure}

\begin{figure}[!ht]
	\centering
	\includegraphics[width=0.9\linewidth]{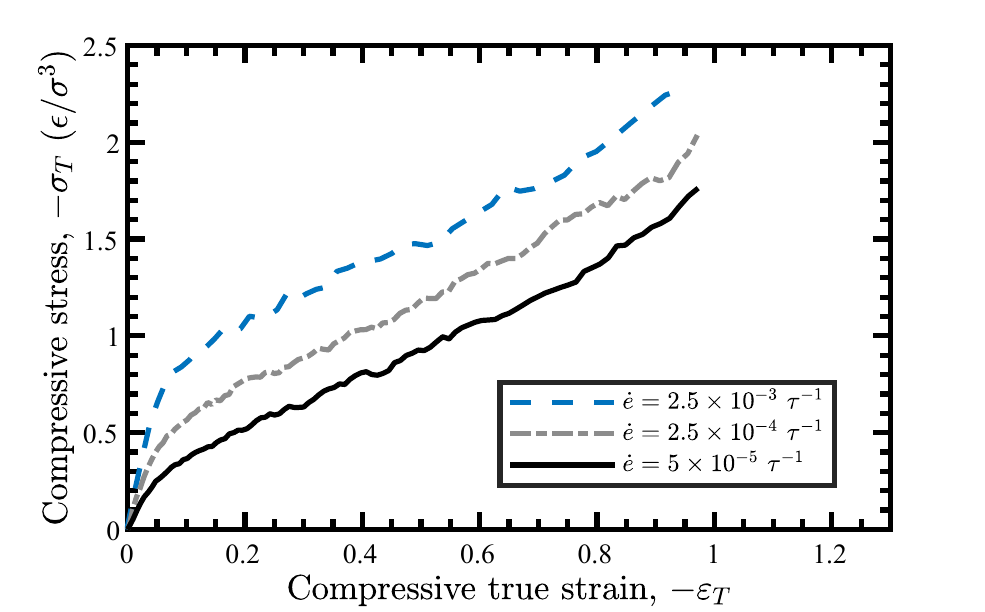}
	\includegraphics[width=0.9\linewidth]{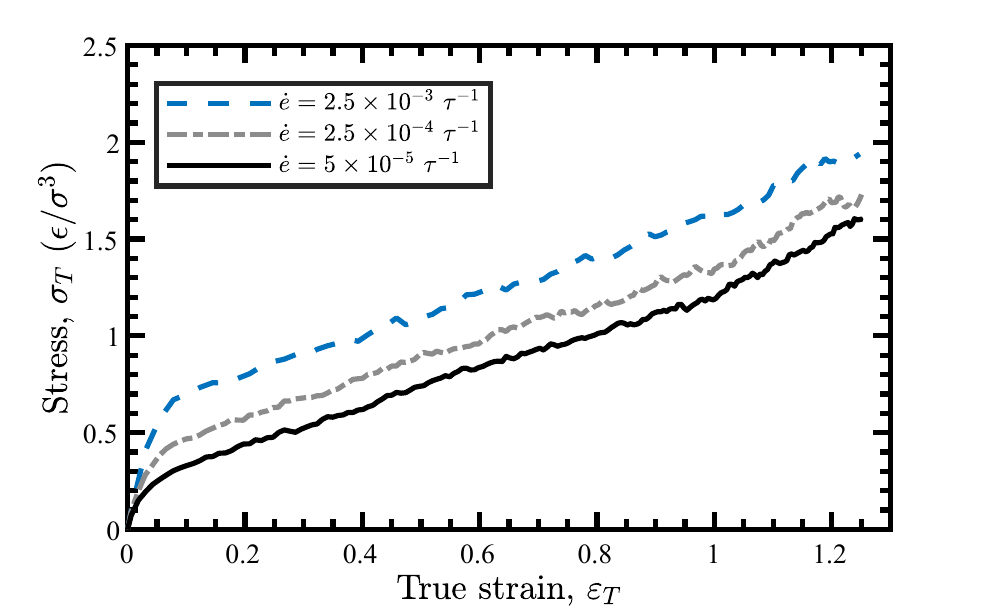}
	\caption{Uniaxial stress-strain curves in compression (top) and tension (bottom) for $k_{\epsilon}=5$.}
	\label{fig5}
\end{figure}
We applied uniaxial strain in tension and compression at three different strain rates to obtain the mechanical response of the RVE. \fref{fig4} compares the resulting stress-strain response for $k_{\epsilon}=2,~3$ and $5$ with the experimental data for polyurea~\cite{sarva2007stress,roland2007high,pathak2008structure}. Stress has been normalized by dividing it with the stress at the maximum simulated true strain ($\varepsilon_T=-0.98$ in compression and $\varepsilon_T=1.2$ in tension). Since stress $\sim$ energy/length$^3$, the normalization helps remove the effect of energy scale ($\epsilon$) and length scale ($\sigma$) and bares the effect of microstructure and time scale. Since strain rates is simulation is expected to be higher than the experimental data, the simulated response in compression conforms to the effect of strain rate. The response in tension however does not conform to the expected strain rate effect. Nevertheless, the close quantitative agreement between the experimental data and the simulated response, specifically corresponding to $k_{\epsilon}=5$, reinforces the validity of the coarse-grained model. Also, the stress-strain response for the three values of $k_{\epsilon}$ are qualitatively similar, though the response becomes stiffer with the increasing value of $k_{\epsilon}$. It also exhibits rate dependence and strain hardening, as expected. Furthermore, the simulated response in tension and compression is not symmetric.

\fref{fig5} shows the effect of loading rate on the response of the RVE for $k_{\epsilon}=5$. Increasing strain rate leads to a stiffer initial response. This is observed both in compressive as well as tensile responses. However, experiments in~\cite{sarva2007stress} show that the initial response in compression is stiffer than in tension but becomes softer after the steep initial rise. While simulated initial response follows the experimental trend, compressive response does not become softer than tensile response after the initial rise. It should be noted however that the maximum strain rate at which experimental tensile response is measured and compared with compressive response is small compared to the expected strain rate in simulation. These deviations from experimental findings likely result from an artificial speed up in microstructual evolution due to coarse-graining as discussed in \sref{microstructure}.

\subsection{Microstructure evolution}
\label{microstructure}
\begin{figure*}
	\centering
	\includegraphics[width=0.75\linewidth]{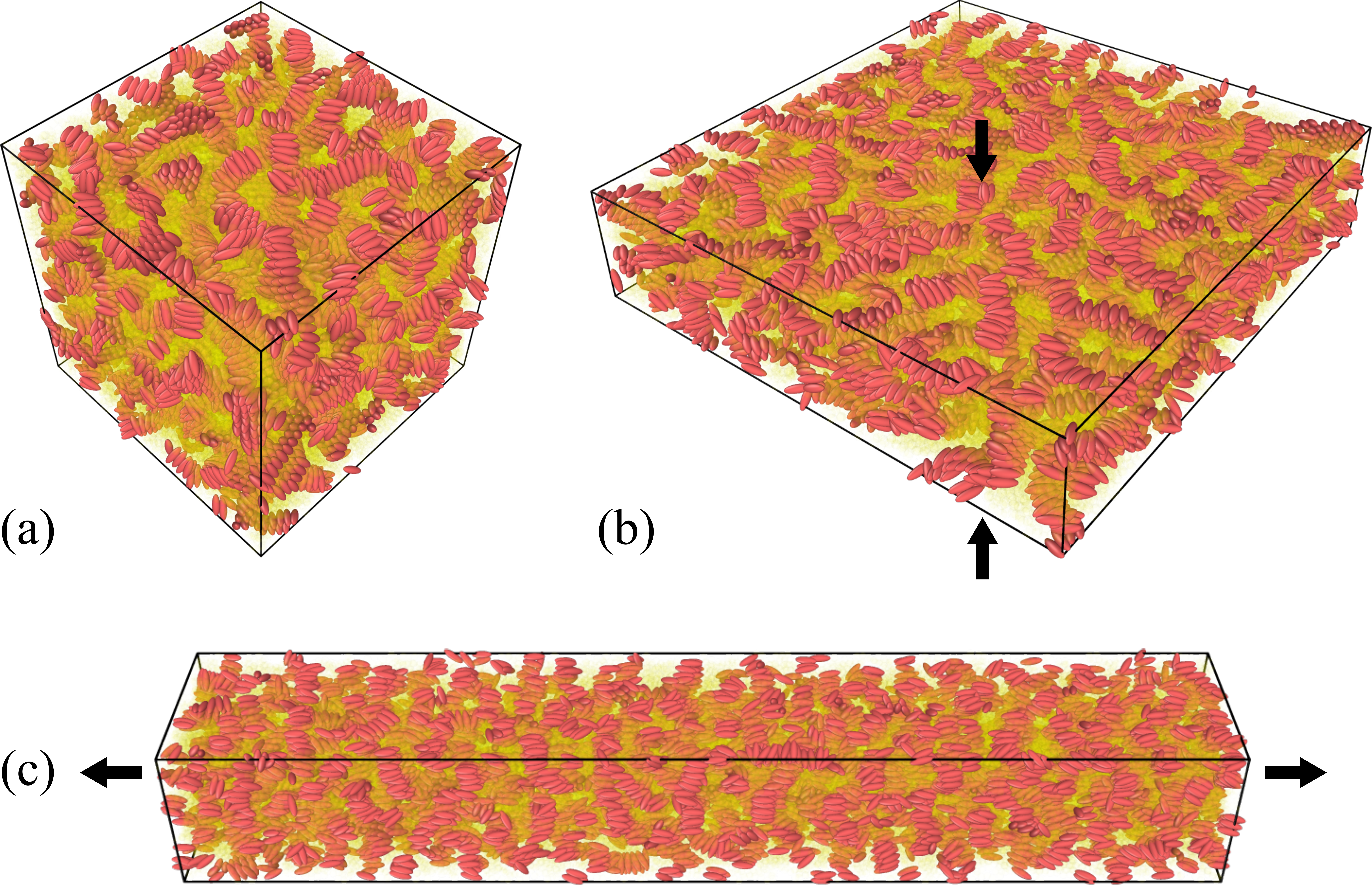}
	\caption{Figure shows RVE in (a) equilibrium, (b) at $\varepsilon_T=-0.98$, and (c) at $\varepsilon_T=1.2$. A translucent soft domain allows the ribbon-shaped hard domain to be exhibited and a fragmented hard domain can be observed in tension.}
	\label{fig6}
\end{figure*}
The initial microstructure in simulation evolves with the applied uniaxial loading. \fref{fig6}(a) shows the initial microstructure in equilibrium at no mechanical loading. The ribbon-shaped hard domain is randomly aligned when no loading is applied. Also, while the neighboring ellipsoids prefer to align in a similar direction, no long-range correlation in alignment is observed.

\begin{figure}[!ht]
	\centering
	\includegraphics[width=0.9\linewidth]{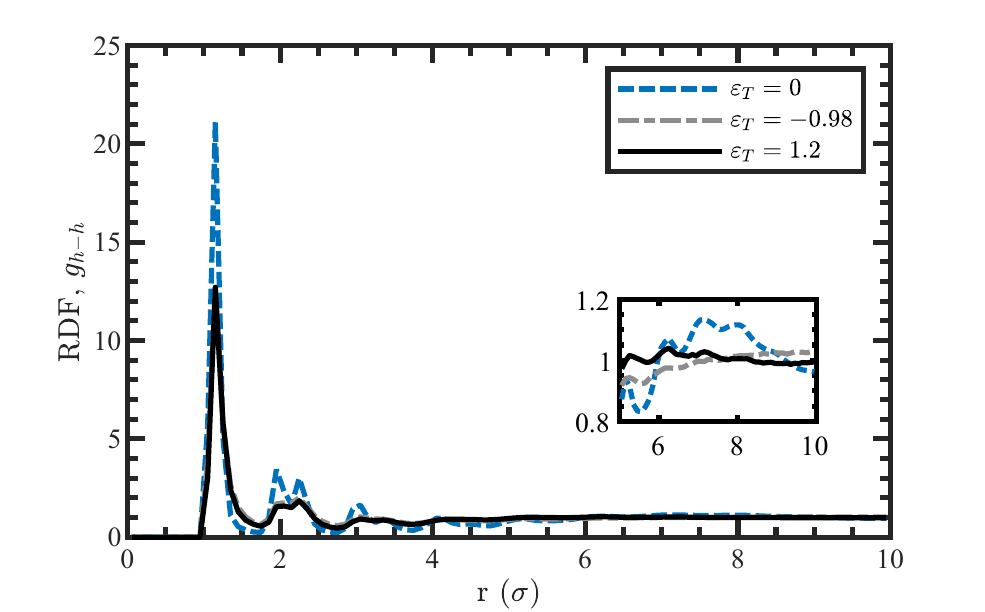}
	\caption{Radial distribution function (RDF) of the hard segments for $k_{\epsilon}=5$.}
	\label{fig7}
\end{figure}

We also performed cluster analysis of ellipsoids by taking the neighbor cut-off to be $1.75\sigma$ at which the first trough in radial distribution function (RDF) is located (see~\fref{fig7}). On defining a hard domain to consist of a minimum of 10(5) ellipsoids, the RVE has two hard domain clusters at no load. Remarkably, it increases to 56(91) clusters at $\varepsilon_T=1.2$, but only 5(11) clusters at $\varepsilon_T=-0.98$. Increasing number of clusters indicates fragmentation of hard domains on the application of a load and a precursor to a plastic deformation mechanism. Moreover, the stark contrast in the number of hard domain clusters in tension and compression points to differing mechanisms of mechanical response in tension and compression. Furthermore, the number of ellipsoidal clusters mixed in the soft domain, taken as clusters with a maximum of 5(1) ellipsoids, is 4(3) at equilibrium. This grows to 180(103) at $\varepsilon_T=1.2$ and 41(28) at $\varepsilon_T=-0.98$. This is indicative of phase mixing due to the applied mechanical load.

RDF of the ellipsoidal particles is plotted in \fref{fig7}. The location of the first peak in RDF does not change with loading, indicating that the mean spacing between neighboring ellipsoids does not change with loading which is in agreement with experimental findings in~\cite{rinaldi2011microstructure,choi2012microstructure}. Furthermore, RDF exhibits order in the hard domain arrangement at no mechanical load, and the spacing between the domains is $\sim 7\sigma$. Tensile loading results in an overall decrease in spacing between domains making it $\sim 6\sigma$ at $\varepsilon_T=1.2$, a trend consistent with experimental results for polyurea~\cite{rinaldi2011microstructure}. Under compressive loading, however, no corresponding peak in RDF is observed up to $r= 10\sigma$.

\begin{figure}[!ht]
	\centering
	\includegraphics[width=0.9\linewidth]{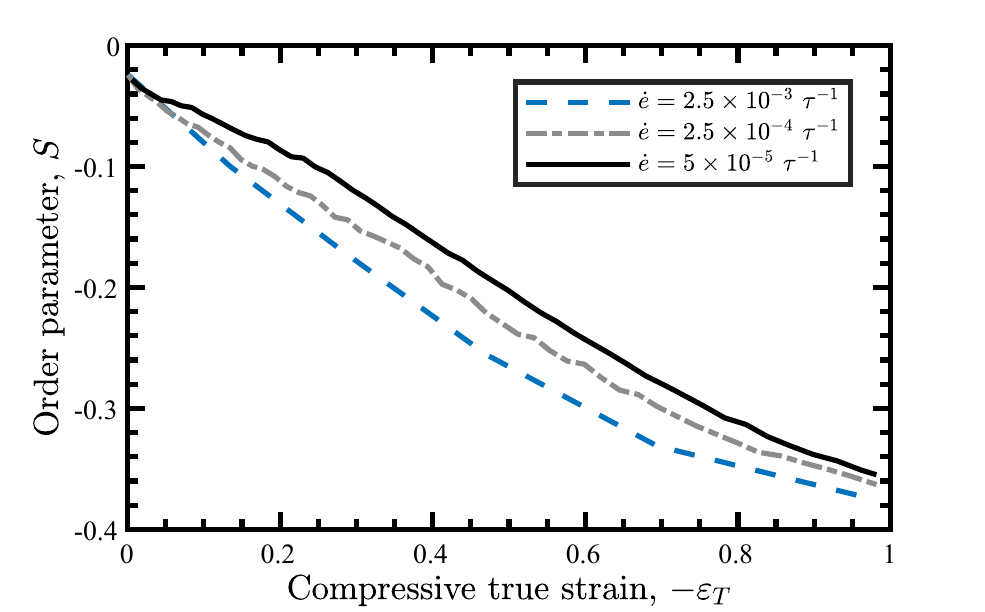}
	\includegraphics[width=0.9\linewidth]{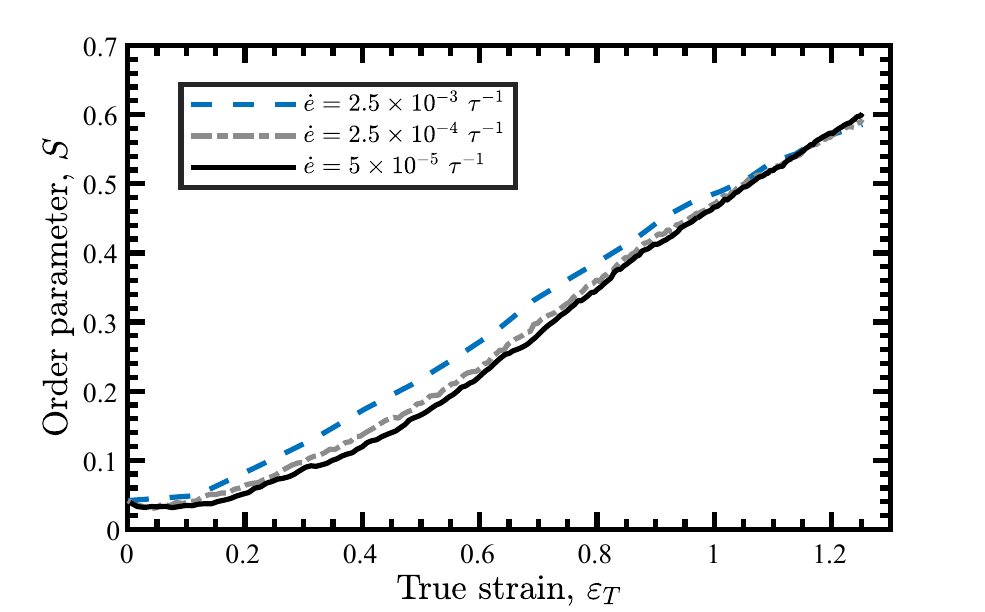}
	\caption{Evolution of order parameter with strain in compression (top) and tension (bottom) for $k_{\epsilon}=5$. The negative value of $S$ in compression indicates that the ellipsoidal particles are aligned in the plane perpendicular to the loading axis.}
	\label{fig8}
\end{figure}
The evolution of the nematic order parameter ($S$) with the applied uniaxial load is plotted in~\fref{fig8}. While ellipsoidal particles exhibit local nematic order at no load as observed in~\fref{fig6}(a), there is no orientational order globally as shown in \fref{fig8}. With increasing compressive stress, $S$ takes a negative value with increasing magnitude. This indicates that ellipsoids align in a plane perpendicular to the loading axis (see \fref{fig6}(b)). In contrast, $S$ increases with increasing tensile loading. This suggests that ellipsoids align with the loading axis (see \fref{fig6}(c)), as observed for the hard segments in polyurea~\cite{rinaldi2011microstructure,choi2012microstructure}. Also, $S$ shows little change for small strains up to $\varepsilon_T\sim 0.2$, indicating that the response is not driven by ellipsoidal alignment at small strains. Furthermore, strain rate has little effect on alignment, as shown in~\fref{fig8}. This is inconsistent with experimental findings for polyurea in~\cite{pathak2008structure}, where alignment is found to be strain rate dependent. This behavior likely results from modeling the hard segments of polyurea as smooth ellipsoidal particles in the simulation, leading to an artificial decrease in time scale associated with the rotational motion of the hard segments~\cite{meinel2020loss}. On the other hand, microprojectile impact tests reveal that thin layers of even glassy polymers exhibit significant mobility and considerable impact energy dissipation in contrast to the bulk of the material, which is brittle~\cite{hyon2018extreme}. In polyurea, not only are glassy strips nanometer thick, they are bonded to flexible segments which can flow, increasing the mobility of the glassy strips~\cite{baglay2015communication}.

\begin{figure}[!ht]
	\centering
	\includegraphics[width=0.9\linewidth]{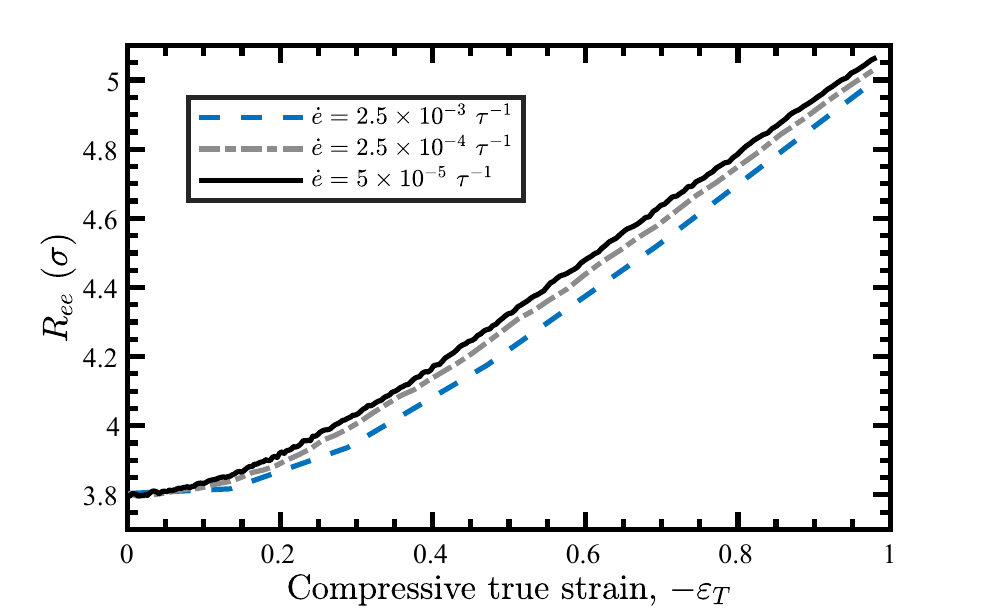}
	\includegraphics[width=0.9\linewidth]{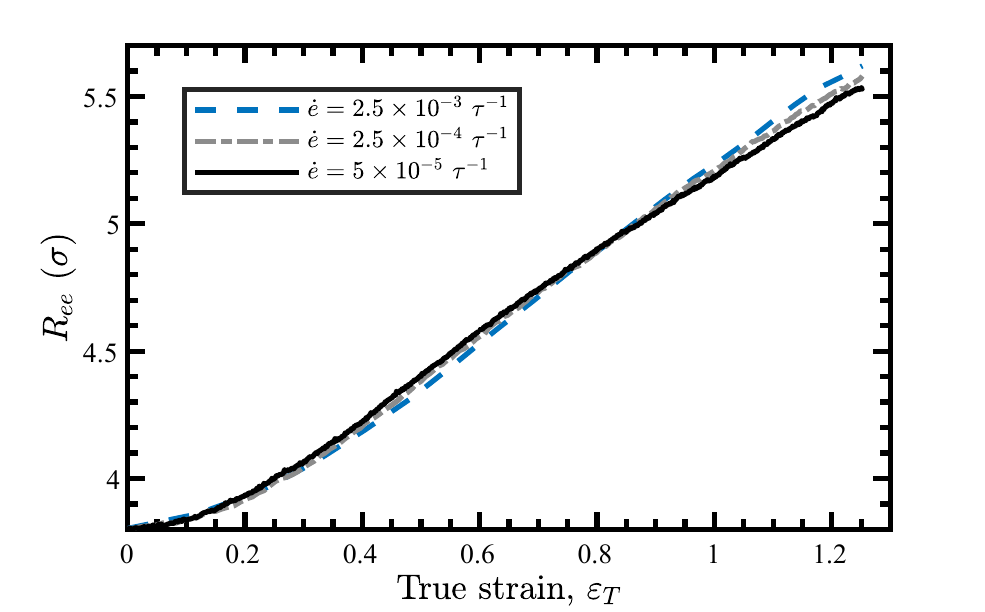}
	\caption{Evolution of end-to-end distance of soft segments connecting two ellipsoidal beads with strain in compression (top) and tension (bottom) for $k_{\epsilon}=5$.}
	\label{fig9}
\end{figure}

\begin{figure}[!ht]
	\centering
	\includegraphics[width=0.9\linewidth]{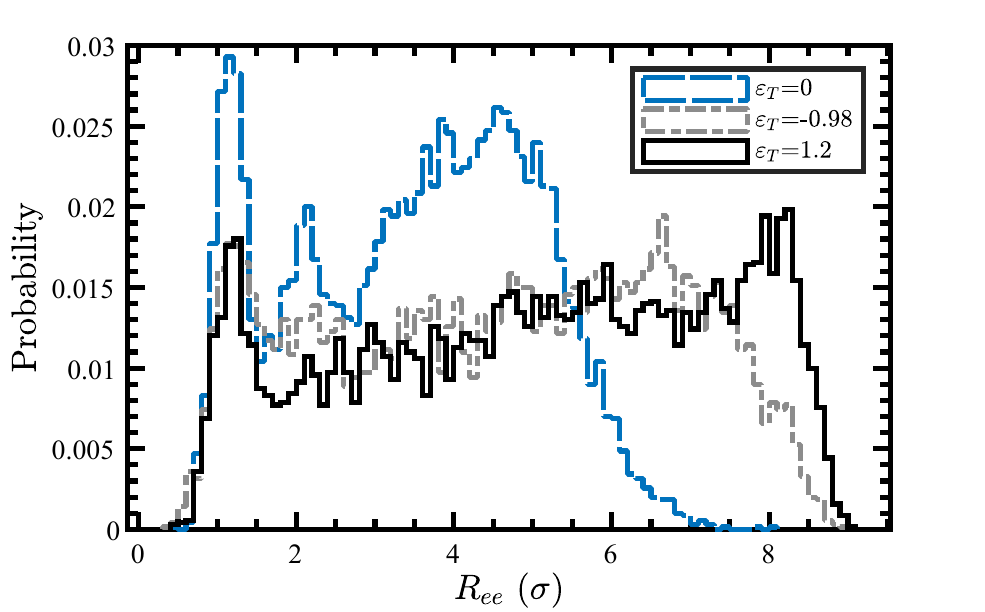}
	\caption{Distribution of end-to-end distance of soft segments connecting two ellipsoidal beads for $k_{\epsilon}=5$.}
	\label{fig10}
\end{figure}
End-to-end distance ($R_{ee}$) of soft segments linking ellipsoids are plotted in \fref{fig9}. It increases in both compression and tension and shows little effect on strain rate, which is expected based on strain rate insensitivity of evolution of the order parameter. We also investigated the distribution of $R_{ee}$ (see~\fref{fig10}), which indicates if they form connecting loops or bridge neighboring ribbons~\cite{zhu2018molecular}. At no load, two sharp peaks and a broad peak are observed. The sharp peaks correspond to the first two peaks in the RDF of the hard segments in~\fref{fig7}. The first peak at $R_{ee}\sim \sigma$ and the second peak at $R_{ee}\sim 2\sigma$ correspond to loop formation, and the third broad peak corresponds to bridge connections. Both tensile and compressive loads lead to the disappearance of the second peak, a decrease in the height of the first peak, and the flattening of the last peak. The disappearance of the second peak and a decrease in the height of the first peak indicates possible partial or complete pulling out of some ellipsoids from the ribbons and their mixing with the flexible domain. This is another plastic deformation mechanism.

\section{Conclusions}
\label{conclusions}
To elucidate the effect of microstructure on the mechanical response of polyurea like hard-soft block copolymers, we have developed a coarse-grained model by representing rigid and anisotropically-shaped hard segments with ellipsoids like in a liquid crystalline polymer. The coarse-grained model successfully produces a microstructure with a ribbon-shaped hard domain akin to the microstructure of polyurea. Remarkably, the normalized uniaxial stress-strain relation in tension as well as in compression for the generic model shows an excellent quantitative agreement with experimental data for polyurea. The model also largely conforms to the experimental findings on the evolution of polyurea microstructure in uniaxial loading. Furthermore, we find that fragmentation of the hard domain ribbons and partial or complete pull-out of hard segments from the ribbons and subsequent phase mixing are possible plastic deformation mechanisms.

This work is confined to developing the generic model and studying its uniaxial mechanical response. Investigation of the dynamic mechanical response of the model remains a work for the future.

\begin{acknowledgments}
We acknowledge the support from the Natural Sciences and Engineering Research Council of Canada (NSERC) through the Discovery Grant under Award Application Number 2016-06114. This research was supported in part through computational resources and services provided by Advanced Research Computing at the University of British Columbia.
\end{acknowledgments}

\bibliography{references}

\end{document}